\documentclass[a4paper,11pt]{article}
\pdfoutput=1 % if your are submitting a pdflatex (i.e. if you have
\usepackage{jcappub} % for details on the use of the package, please
\usepackage{enumerate}
\usepackage[T1]{fontenc} % if needed
\usepackage{caption}
\usepackage{subcaption}
\usepackage{empheq}
\usepackage{hyperref}
\usepackage{natbib}
\usepackage{subfiles}
\usepackage{aas_macros}
\usepackage{amsmath}
\usepackage{amssymb}
\usepackage[section]{placeins}
\usepackage{graphicx}
\usepackage{booktabs}
\usepackage[normalem]{ulem}
\usepackage{aas_macros}
\useunder{\uline}{\ul}{}
\newcommand{\unit}[1]{\ensuremath{\, \mathrm{#1}}}
\newcommand{\vbar}{\bar{v}}
\newcommand{\vesc}{v_\text{esc}}
\newcommand{\Rstar}{R_{\star}}
\newcommand{\Mstar}{M_{\star}}
\newcommand{\be}{\begin{equation}}
\newcommand{\ee}{\end{equation}}
\newcommand{\pn}{p_N(\tau)}

\newcommand{\vn}{v_{N}}
\newcommand{\expmess}{\exp{\left(-\frac{3(\vn^2-\vesc^2)}{2\vbar^2}\right)}}

\newcommand{\avg}[1]{\langle #1 \rangle}

\newcommand{\soneinf}{\sum_{N=1}^{\infty}}

\newcommand{\GeV}{\unit{GeV}}

\newcommand{\percc}{\unit{cm}^{-3}}

\title{Closed-form expressions for  multiscatter Dark Matter capture rates}

\author{Cosmin Ilie}

\affiliation{Colgate University,\\13 Oak Drive, Hamilton, NY}

\emailAdd{cilie@colgate.edu}

\abstract{Any astrophysical object can, in principle, serve as a probe of the interaction between Dark Matter and regular, baryonic matter. This method is based on the potential observable consequences annihilations of captured Dark Matter has on the surface temperature of the object itself. In a series of previous papers we  developed and validated simple analytic approximations for the total capture rates of Dark Matter (DM) valid in four distinct regions of the DM-nucleon scattering cross section ($\sigma$) vs. DM particle mass ($m_X$) parameter space. In this work we summarize those previous results and extend them significantly, by deriving  a completely general, closed form solution for the total capture rate of Dark Matter in the multiscatter regime. Moreover, we demonstrate the existence of a region in the  $\sigma$ vs. $m_X$ parameter space where the constraining power of any astrophysical object heated by annihilations of captured DM is lost. This corresponds to a maximal temperature ($T_{crit}$) any astrophysical object can have, such that it can still serve as a DM probe. Any object with observed temperature $T_{obs}>T_{crit}$ loses its DM constraining power. We provide analytic formulae that can be used to estimate $T_{crit}$ for any object.}

\begin{document}
\maketitle
\flushbottom

\section{Introduction}
\label{sec:intro}

One of the most profound problems in modern physics is the that of the nature of Dark Matter (DM). The origin of the term can be traced back to the work of Fritz Zwicky in 1933~\cite{Zwicky:1933} who predicted the existence of a non-luminous component of matter in the Coma Cluster of galaxies to explain the measured velocity dispersion. The idea of DM remained dormant in the literature until the 1970s when Rubin and Ford inferred its existence on galactic scales by measuring the rotational speed of H2 gas surrounding the Andromeda Galaxy~\cite{Rubin:1970}. Since then, a plethora of evidence has emerged supporting the DM hypothesis, including gravitational lensing~\citep[e.g.][]{Tyson_1998}, the bullet cluster~\citep[e.g.][]{Clowe_2006}, and the Cosmic Microwave Background~\citep{Komatsu:2009,Komatsu:2011,Ade:2015,Aghanim:2018}, which predicts that DM composes $\sim 85\%$ of the mass-budget of the universe.~Despite the clear indicators of DM's existence through gravitational phenomena, the exact nature of DM has remained elusive despite the combined efforts from theorists and experimentalists. Various candidates for the form of DM have emerged since its existence was inferred. These include massive compact halo objects (MACHOs) and various forms of particle DM, such as Axions and Weakly Interacting Massive Particles~(WIMPs), to name a few. Massive compact halo objects have largely been ruled out as dark matter candidates from experimental data and theoretical arguments while Axions, WIMPs, and other particle DM candidates have been heavily constrained by experiments like ADMX~\cite{Braine_2020}, XENON1T~\cite{Aprile:2019}, and PICO60~\cite{Amole:2019fdf}, which are designed to measure the effect of DM particles on detectors deep underground. For a review of the status of Dark Matter research, both theoretical and experimental, see Ref.~\cite{Freese:2017dm}.

Despite our lack of understanding DM's true form, a clear picture has emerged for the role of DM in the structure of our universe in the standard cosmological model, $\Lambda$CDM. Overdensities of DM form in the early universe after a period of rapid expansion, known as cosmic inflation, which amplify quantum fluctuations in the early universe~\citep[e.g.][]{liddle_lyth_2000}. These dense regions of DM, known as halos, provide the gravitational impetus for baryonic matter to collapse and form the first stars~\citep[e.g.][]{Bromm:2003}. Through mergers, these DM halos grow, providing the seeds for galaxy formation~\citep[e.g.][]{Bromm:2009}. Simulations of DM in the early universe confirm this picture and also posit the existence of a filament-like structure connecting the DM halos~\citep[e.g.][]{Wang_2020}. This hierarchical structure formation predicts that most, if not all, galaxies are formed in DM halos that have a rich sub-structure. 

Dark Matter capture is a mechanism for accumulating particle DM within a compact object~\citep[e.g.][]{Gould:1987,Bramante:2017,Ilie:2020Comment,Dasgupta:2019juq}. DM particles incident on an object first become accelerated by the object's gravitational field, then, as they transit the object, may collide with its constituents, losing kinetic energy. Provided the particle loses enough energy, it will become gravitationally bound to the object. From there, a number of processes may occur that leaves a DM footprint on the object, including DM-DM annihilation, DM-baryon annihilation, and kinetic heat transfer. Capture, and its observational effects, have long been studied and are of serious interest in the search for DM since many physical situations consist of astrophysical objects within dense DM environments, which intuitively produces higher rates of capture. A non-exhaustive list of previous literature on the observational effects of captured dark matter include the study of neutron stars~\citep[e.g.][]{Bramante:2017, Ilie:2020Comment,Baryakhtar:2017,Garani:2020}, Population III stars~\citep[e.g.][]{Ilie:2019,Ilie:2020PopIII,Ilie:2021mcms}, white dwarfs~\citep[e.g.][]{Dasgupta:2019juq,Horowitz:2020axx}, and exoplanets~\citep[e.g.][]{Leane:2020wob}. 

In Refs.~\cite{Ilie:2020Comment,Ilie:2020PopIII} we developed simple analytic approximations for the capture rates of DM particles valid in the regime when the escape velocity ($\vesc$) is much greater than the average DM velocity ($\vbar$). A natural picture emerged, where DM capture rates take simple analytic forms (see Sec.~\ref{ssec:Preamble} for a summary) in four different regions in the $\sigma$ (DM-SM interaction cross section) vs $m_X$ (DM particle mass)  parameter space. Combined, all those four regions cover the entire $\sigma$ vs $m_X$ plane. The opposite regime ($\vesc\ll\vbar$) was analyzed by us in Ref.~\cite{Ilie:2023exo}. We uncovered a new region, in addition to the four previously identified (where the simple analytic approximations still hold). In this newly identified region we are, as of yet, unable to generate simple analytic approximations. Motivated by this limitation, in this work we report our findings of a general analytic solution for the total multiscatter capture rate, valid independent of the location in the $\sigma$ vs. $m_X$ plane (See sec.~\ref{ssec:SCMS}) and independent of the capturing object (i.e. no hierarchy between $\vesc$ and $\vbar$ assumed). Albeit somewhat cumbersome, we find this analytic solution extremely useful, as it avoids laborious numerical computations of the total DM capture rates, which, sometimes are prone to numerical convergence issues.

\section{Brief Review of (Multiscatter) Dark Matter Capture formalism}\label{sec:Review}

The problem of Dark Matter capture by celestial bodies, its reverse process (evaporation), and possible observable effects of those processes was first studied in the 80's and early 90's in a series of seminal papers by pioneers in the field such as Spergel~\& Press~\cite{Spergel:1985,Press:1985}, Gould~\cite{Gould:1987,Gould:1987resonant,Gould:1987evap,Gould:1989,Gould:1990_WimpConduction,Gould:1992ApJ}, among others. For more recent developments see Refs.~\cite{Bramante:2017,Garani:2017,Dasgupta:2019juq,Bell:2020NSSINS,Bell:2020b,Garani:2022,Ilie:2023exo,Acevedo:2023owd}, to name a few. 

Whenever a DM particle crosses an astrophysical object, it can lose energy, via collisions with the baryons inside the object, and become gravitationally trapped (i.e. captured) whenever its velocity becomes less than the escape velocity at the surface of the star. The total capture rate can be represented as a sum of independent (partial) capture rates:
\be\label{eq:Ctot}
C_{tot}=\soneinf C_N,
\ee
where $C_N$ represents the capture rate after exactly $N$ collisions and is given by~\citep{Bramante:2017}: 

\be\label{eq:CNIntegralForm}
C_N=\underbrace{\pi \Rstar^2}_\textrm{capture area}\times \,\underbrace{n_X \int_0^{\infty} \dfrac{f(u)du}{u}\,(u^2+v_{ esc}^2)}_\textrm{DM flux}\times \, \underbrace{p_{ N}(\tau)}_\textrm{prob. for $N$ collisions}\times \, \underbrace{g_{ N}(u)}_\textrm{prob. of capture},
\ee
where $\Rstar$ is the radius of the object, $u$ is the speed of the DM particle far from the object, $f(u)$ is the DM speed distribution in the medium, $\vesc$ is the surface escape velocity of the object, $\tau \equiv 2 \Rstar \sigma n_T$ is the optical depth,\footnote{Physically, this represents the average number of scatters a DM particle would undergo if it traversed the diameter of the object.} with $n_T$ representing the average density of scattering targets in side the object. For non-relativistic (cold) DM we have: $n_X=\frac{\rho_X}{m_X}$. The probability of $N$ collisions between DM and nuclei inside the star ($p_N(\tau)$) has the following closed form~\citep{Ilie:2019}:
\be\label{eq:pN}
\pn=\frac{2}{\tau^2}\left(N+1-\frac{\Gamma(N+2,\tau)}{N!}\right), 
\ee
where $\Gamma(a,b)$ is the incomplete gamma function. For the probability that a DM particle is slowed down below $\vesc$ by exactly N collisions we assume, following~\cite{Bramante:2017}:   
\be\label{eq:gN}
g_N(u)=\Theta(u_{max;N}-u),
\ee
where $\Theta(x)$ is the Heaviside step function. Throughout, we denote by $$u_{max;N}=\vesc\left[(1-\beta_+/2)^{-N}-1\right]^{1/2}$$ the maximum value of the velocity a DM particle can have, far from the star, such that it will be slowed down below the escape velocity after $N$ collisions. Here $\beta_+\equiv 4mm_X/(m+m_X)^2$, with $m$ being the mass of the target nucleus. For object at rest with respect to the DM halo (i.e. such as those located close to the center)  the DM velocities (as seen by the capturing object) follow a usual Maxwell-Boltzmann distribution ($f_{MB}(u)$), for which the only controlling parameter is the velocity dispersion ($\vbar$).  Under this assumption the capture rate after exactly $N$ scatters is~\cite{Bramante:2017,Ilie:2020Comment}:
\be\label{eq:CN}
C_{N}=\frac{1}{3}\pi R_{\star}^{2} p_{N}(\tau) \sqrt{\frac{6}{\pi}}\frac{n_{X}}{ \bar{v}}\left(\left(2 \bar{v}^{2}+3 v_{e s c}^{2}\right)-\left(2 \bar{v}^{2}+3 v_{N}^{2}\right) \exp \left(-\frac{3\left(v_{N}^{2}-v_{e s c}^{2}\right)}{2 \bar{v}^{2}}\right)\right),
\ee
where $v_N=\vesc(1-\avg{z}\beta_+)^{-N/2}$ is the velocity of DM after $N$ scatters, and $\langle z\rangle$ is a number between zero and one, that accounts for the scattering angle.\footnote{After elastically colliding with a nucleus, the particle with $E_0$ kinetic energy will lose an amount of energy:
$\Delta E = z \beta_+ E_0$, where $z$ is related the scattering angle in the center of mass frame by $z=\sin^2(\theta_{\text{CM}}/2)$\citep{Dasgupta:2019juq}.} Whenever the differential dark matter-nuclear cross section is independent of scattering angle one can usually approximate $\langle z\rangle\approx 1/2$. Moreover, note that $\beta_+\leq 1$, attaining the maximum value whenever $m_X=m$. Whenever $\avg{z}\beta_+N\ll1$, one can show that $C_N$ can be recast in the following form~\citep{Ilie:2023exo}:
\be\label{eq:CN-newaprx}
C_{N}=\frac{1}{3}\pi R_{\star}^{2} p_{N}(\tau) \sqrt{\frac{6}{\pi}}\frac{n_{X}}{ \bar{v}}\left(2\vbar^2+3\vesc^2\right)\left[1-\left(1+\frac{3\avg{z}\beta_+\vesc^2}{2\vbar^2+3\vesc^2}N\right)e^{-kN}\right],
\ee
with the parameter k defined as:
\be\label{eq:k-newdef}
k\equiv\frac{3\vesc^2}{2\vbar^2}\avg{z}\beta_+\equiv\alpha\beta_+,
\ee
where in the last definition we introduced the following useful notation: $\alpha\equiv\frac{3\vesc^2}{2\vbar^2}\avg{z}$. Whenever there is a large hierarchy between $\vesc$ and $\vbar$ (i.e. $\vesc \gg \vbar$ or $\vesc\ll\vbar$) and, similarly, a large hierarchy between $m_X$ and $m$, the $C_N$ in Eq.~\ref{eq:CN-newaprx} can be further simplified, since now $\beta_+\simeq4\frac{\min(m_X,m)}{\max(m_X,m)}$. For instance in Ref.~\cite{Bramante:2017} one is interested in capture of very heavy DM (i.e. $\beta_+\simeq4\frac{m}{m_X}$) by compact objects (i.e. $\vesc\gg\vbar$). Furthermore, assuming $\avg{z}=1/2$ one gets:  
\be\label{eq:CN-BDM23-new}
C_{N} = \pi R_{\star}^{2} p_{N}(\tau) \sqrt{\frac{6}{\pi}}\frac{n_{X}\vesc^2}{ \bar{v}}\left(1-\left(1+\frac{2 A_{N}^{2} \bar{v}^{2}}{3v_{esc}^{2}}\right) e^{-A_{N}^{2}}\right), \text{where} \  A_{N}^{2}=\frac{3N\vesc^2 }{\vbar^2}\frac{\min(m_X;m)}{\max(m_X;m)}.
\ee
This last result is the same with Eq.~23 of~\cite{Bramante:2017}, if one replaces $\vesc^2$ with the Newtonian $2G\Mstar/\Rstar$. Note, however, that for Neutron Stars one should instead use the relativistic version: $\vesc=\sqrt{2 \left[1-\left(1-\frac{2G\Mstar}{\Rstar c^2}\right)^{1/2}\right]}$. Moreover, general relativistic effects lead to an enhancement of the capture rate coefficients $C_N$ in a manner that can be approximated by the following prescription~\cite{Bramante:2017}:
\be\label{eq:CN-rescaledGR}
C_N\to \frac{C_N}{1-2G\Mstar/\Rstar}.
\ee

In what follows we discuss several limitations of the multiscatter DM capture formalism summarized above. First, we point out that the expression listed for $C_N$ in Eqn.~\ref{eq:CN}, and its subsequent approximation (e.g. Eq.~\ref{eq:CN-BDM23-new}) are valid under the assumption that the capturing object has zero relative velocity with respect of the DM halo restframe, i.e. it is located very close to the center of the DM halo. For objects with a non zero velocity relative to the DM halo rest frame the DM capture rates are suppressed, as first pointed out in Ref.~\cite{Gould:1987resonant}. In Ref.~\cite{Ilie:2021vel} we worked out an analytic prescription for total capture rate suppression factor, as a function of the relative velocity. Throughout this work, as done in most of the DM capture literature, we will assume that the capture rate is not suppressed. However, if one wants to apply the analytic results we develop here for the case of objects with non zero relative velocity w.r.t. the DM halo, then one could do so by including the overall suppression factor found in Ref.~\cite{Ilie:2021vel}. Additionally, for many objects there is the extra complication of capture via multiple scatterings off of different targets. With this in mind, in Ref.~\cite{Ilie:2021mcms} we developed the multi-component multi-scatter formalism and applied it to investigate the role of the Helium nuclei on the efficiency of Pop~III stars as DM probes. 

Analytic approximations of the total DM capture rates ($C_{tot}$) have been obtained in Ref.~\cite{Ilie:2020PopIII} (valid whenever $\vesc\gg\vbar$) and in Ref.~\cite{Ilie:2023exo} (valid whenever $\vesc\ll1$ and only covering a subset of the $\sigma$ vs $m_X$ parameter space).  In the next section we review those results, and then proceed to derive a general closed form solution for the total capture rate, valid anywhere in the $\sigma$ vs $m_X$ plane, and for any capturing object (i.e. no hierarchy between $\vesc$ and $\vbar$ is assumed). 

\section{Analytic closed-forms of DM Capture rates and their approximations}\label{sec:Analytic}

In this section we present closed form analytic formulae for DM capture rates. We begin in Sec.~\ref{ssec:Preamble}, with a summary of our previous findings: analytic approximations for the two limiting cases: $\vesc\gg\vbar$ (Ref.~\cite{Ilie:2020PopIII}) and $\vesc\ll\vbar$ (Ref.~\cite{Ilie:2023exo}), and the emergence of four regions in the $\sigma$ vs. $m_X$ parameter space where said approximations are valid. Motivated by the surprising discovery of a fifth region, emerging the case of objects with $\vesc\ll\vbar$ (e.g. exoplanets), for which no closed form approximation for the total capture rate is found, we derive in Sec.~\ref{ssec:SCMS} a completely general analytic closed form of the capture rate, valid independent of capturing object or location in the $\sigma$ vs. $m_X$ plane.  

\subsection{Summary of our Previous Results}\label{ssec:Preamble}

In Refs.~\cite{Ilie:2020Comment,Ilie:2020PopIII} we developed and validated analytic expressions for the total capture rate in the limiting regime of $\vesc \gg \vbar$ and for the commonly used case of $\langle z \rangle \sim \frac{1}{2}$. Both of those assumptions are valid for Pop~III stars (the objects of interest in Ref.~\cite{Ilie:2020PopIII}) or for Neutron Stars~\citep{Bramante:2017}. The result, nicely summarized by Fig.~4 of Ref.~\cite{Ilie:2020PopIII}, is a set of analytic expressions for four different regions of $\sigma-m_X$ parameter space, divided according to the optical depth, $\tau$, and the dimensionless kinematic variable $k$. In Ref.~\cite{Ilie:2023exo} a more general analysis was performed, leading to the derivation of approximations for the capture rates regardless of $\vbar-\vesc$ hierarchy, and for all possible values of $\langle z\rangle$. This was motivated by the fact that many of the exoplanet models considered in Ref.~\cite{Ilie:2023exo} straddle the boundary of $\vbar\sim \vesc$, so it was necessary to have a clear picture of the analytic capture rates in this more general scenario.

\begin{figure}[!thb]
    \centering
    \includegraphics[width=1\textwidth]{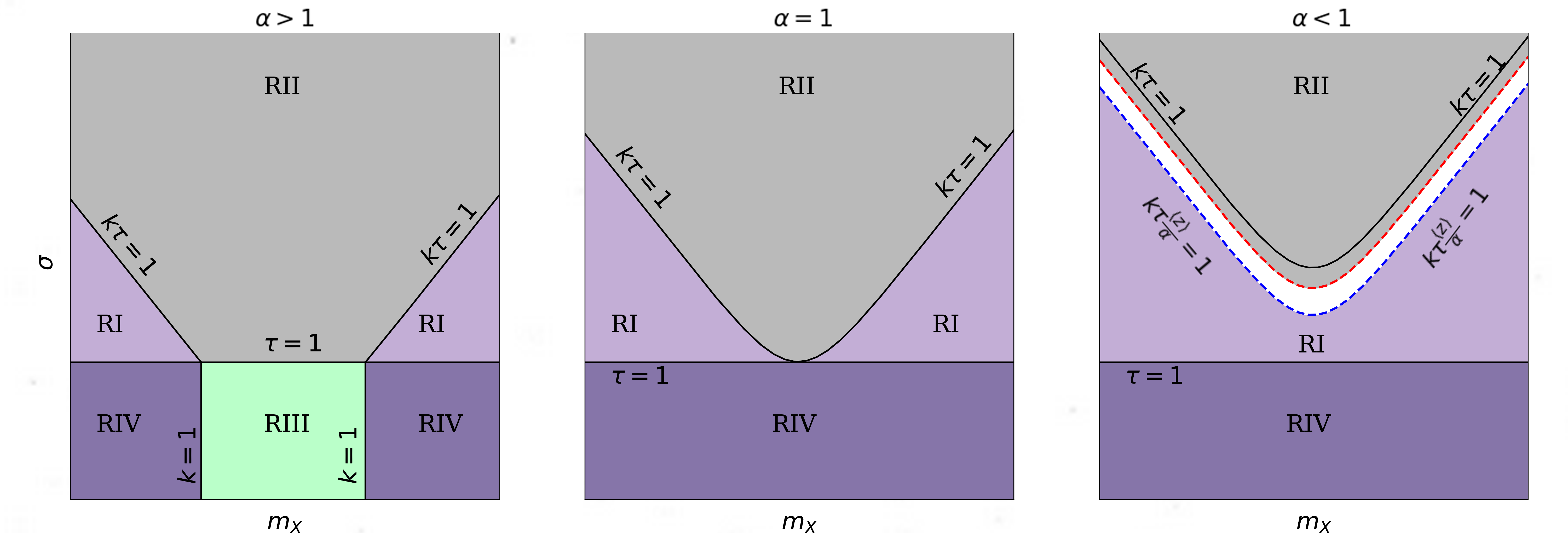}
    \caption{Figure reproduced from Ref.~\cite{Ilie:2023exo}. Schematic depiction of $\sigma-m_X$ parameter space indicating the four regions (RI-RIV) associated with the analytic capture rates presented in Eqns.~(\ref{eq:CtotRI})-(\ref{eq:CtotRIV}). The left panel ($\alpha >1$) describes the results of Ref.~\cite{Ilie:2020PopIII}. Note that for any given $m_X$, the saturation of the capture rate to its geometric limit happens at a finite and well defined value of $\sigma$, i.e. the boundary of RII.  When $\alpha \leq 1$ (central and right panels) RIII will be erased, since now there is no solution for $k=1$. For the case of $\alpha < 1$ (right panel) we furthermore point out several important facts. First, the boundary of RII is no longer given by $k\tau=1$, as was the case for $\alpha\geq 1$ (left and central panel). This boundary (dashed red line) is now pushed to lower values of $\sigma$, i.e. the capture rates saturate faster. Moreover, the boundary of RI is now given by $k\tau\frac{\langle z\rangle}{\alpha}=1$. As a result, RI and RII are now decoupled, and a new region (white gap) emerges, where no analytic approximations for the total capture rates were found. Motivated by this, in Sec.~\ref{ssec:SCMS} we derive a completely general closed form for $C_{tot}$, valid independent of $\alpha$ or location in the $\sigma-m_X$ parameter space.}
    \label{fig:Regions}
\end{figure}

Depending on the value of $\alpha$\footnote{$\alpha\equiv\frac{3\vesc^2}{2\vbar^2}\avg{z}$, as defined in Eq.~\ref{eq:k-newdef}.} compared to unity, two similar (although slightly different) pictures emerge. The case of $\alpha \gtrsim 1$ was addressed in Ref.~\cite{Ilie:2020PopIII} where we found that the parameter space naturally divided into 4 regions (see left panel of Fig.~\ref{fig:Regions}). In each of those regions (labeled I-IV) the total capture rates take a particularly simple form (for a derivation see Appendix A of Ref.~\cite{Ilie:2020PopIII}):
\begin{eqnarray}
     C_{tot}^I &=& 2 A k\tau \vesc^2,\label{eq:CtotRI}\\
     C_{tot}^{II} &=& A \left(2 \vbar^2 + 3\vesc^2\right),\label{eq:CtotRII}\\
     C_{tot}^{III} &=& 2 A \tau \vesc^2,\label{eq:CtotRIII}\\
     C_{tot}^{IV} &=& 2 A k\tau \vesc^2, \label{eq:CtotRIV}
\end{eqnarray}
where $A\equiv \frac{1}{3}\pi \Rstar^2 \sqrt{\frac{6}{\pi}} \frac{n_X}{\vbar}$. Note that  $C_{tot}^{II}$ is the geometric limit of the capture rate, i.e. the total flux crossing an object. 
As shown in Ref.~\cite{Ilie:2023exo}, the expressions above are also valid in the $\alpha\lesssim 1$ regime. The difference, with respect to the $\alpha\gtrsim 1$ case is the erasure of RIII, and the emergence of a new region separating RI and RII, as one can see from Fig.~\ref{fig:Regions}.  Therefore, the four regions found in Refs.~\cite{Ilie:2020PopIII,Ilie:2023exo} completely describe the analytic capture rates in most of the $\sigma-m_X$ parameter space. In fact, for the case of $\alpha\geq1$, we find that RI-RIV fully cover the entire $\sigma$ vs $m_X$ plane, whereas for $\alpha<1$ (e.g. light capturing objects, such as exoplanets) a new gap regions emerges, between RI and RII (see Fig.~\ref{fig:Regions}). In this newly identified region, represented by a white strip in right panel of Fig.~\ref{fig:Regions}, we cannot find a closed form analytic approximation for $C_{tot}$. In fact, even its upper boundary (red dashed line in Fig.~\ref{fig:Regions}) couldn't be analytically determined. Motivated by those two limitations, in Sec.~\ref{ssec:SCMS} we derive a general closed form analytic solution to $C_{tot}$, valid independent of capturing object or location in $\sigma-m_X$ plane. 

In Fig.~\ref{fig:Regions} we plot the schematic partition of the $\sigma$ vs. $m_X$ log-log parameter space in the four regions identified by us in Refs.~\cite{Ilie:2020PopIII,Ilie:2023exo}. Below we discuss in more detail the four regions found, and some implications. In the single scatter regime (i.e. $\tau\leq1$)we find that there are two distinct cases, depending on the value of $\alpha$. Whenever $\alpha>1$ we have two distinct regions in that regime: RIII and RIV, separated by transitions to $k\sim 1$ lines (left panel of Fig.~\ref{fig:Regions}). As $k$ is insensitive to $\sigma$, those lines are vertical lines corresponding to the two solutions of the quadratic (in $m_X$) equation: $k\equiv\alpha\beta_+(m,m_X)=1$.  In the limit of $\alpha\gg1$ those two solutions have a particularly simple form: $m_X^{L}\simeq\frac{m}{4\alpha}$ and $m_X^{R}\simeq4\alpha m$ (we use L for left and R for right). As such, in a log $m_X$ space $k=1$ lines would appear as symmetric with respect to the mass of the target particle ($m$). As one decreases $\alpha$, RIII becomes more and more squeezed, and once $\alpha$ becomes less than unity this region is completely erased, as there will be no solutions now for $k=1$. In the multiscatter regime, and for $\alpha\geq 1$, the two regions (RI-II) are separated by the the solutions in the $\sigma$ vs $m_X$ parameter space of the $k\tau=1$ condition. This can be viewed as a quadratic equation (in $m_X$): $\alpha\tau\beta_+=1$  Moreover, the $k\tau=1$ boundary is symmetric with respect to $m_X=m$, as one can easily show. In the case of $\alpha<1$ we find that RI and RII are decoupled, as already mentioned. This feature is due to the non-trivial convergence properties of the $C_{tot}=\Sigma C_N$ series whenever $\alpha<1$. 

Below we discuss some of the implications of the emergence of the four regions picture presented schematically in Fig.~\ref{fig:Regions} and validated numerically in Refs.~\cite{Ilie:2020PopIII,Ilie:2023exo}. First of all, we find, as expected, that the total capture rate ($C_{tot}$) in the multiscatter regime ($\tau\gg 1$) will invariably saturate to what is called the ``geometric'' limit, i.e. the case when the entire DM flux crossing a star is captured. This can be seen explicitly from the $\tau$ independence of $C_{tot}$ in RII. The fact that a capture rate saturates to the flux is definitely not a new result. The novelty, at least to our knowledge, is the possibility of the extraction of the lower boundary of RII in the $\sigma$ vs. $m_X$ parameter space. For $\alpha>1$ this boundary is given by the $k\tau=1$ and $\tau=1$ lines (see left panel of Fig.~\ref{fig:Regions}). For $\alpha=1$ the $k\tau=1$ line becomes the boundary of RII (middle panel of Fig.~\ref{fig:Regions}). In the case of $\alpha<1$ the boundary is now below the $k\tau=1$ line, but above the $k\tau\avg{z}/\alpha$ line. Sometimes in the literature there is a misconception that the geometric capture rate is attained at ``infinite'' $\sigma$. However, as it should be clear by now, this is not the case. In fact for $\alpha>1$, and for $m_X\sim m$, the saturation happens as soon as one crosses into the multiscatter regime (i.e. as soon as $\tau\sim 1$). Moreover, the existence of such a region which saturates $C_{tot}$, at finite $\sigma$, has immediate consequences for the literature regarding the use of astrophysical objects as DM probes, as pointed out in Refs.~\cite{Ilie:2020Comment,Ilie:2023exo}. Namely, for every probe that uses excess surface temperature due to internal DM heating, we find a critical temperature ($T_{crit}$) above which said probes lose all constraining power.  For completeness, in Sec.~\ref{sec:Implications} we review this aspect.

 While the existence of the RII (where $C_{tot}$ saturates) is to be expected, one of the most intriguing findings of our work is the fact that the analytic form of the capture rates in another region of the multiscatter capture regime (RI) is precisely the same as the analytic form in RIV (single scattering and $k\ll1$), as one can see from comparing Eqns.~\ref{eq:CtotRI} and~\ref{eq:CtotRIV}. It is as if one can simply forget about the multiscatter process and pay no price, when transitioning between regions RIV (single scattering) and RI (multiscattering). 
 
 In summary, we presented simple approximations of the total capture rates (see Eqns.~\ref{eq:CtotRI}-\ref{eq:CtotRIV}) that hold in the four regions identified by us (see Fig.~\ref{fig:Regions}). The reminder of this paper is organized as follows: in Sec.~\ref{ssec:SCMS} we derive a general closed form formula for the total capture rate, that is valid for all capturing objects (i.e. independent of $\alpha$ or $\avg{z}$) and location in the $\sigma-m_X$ parameter space.  We end with Sec.~\ref{sec:Implications}, where we discuss the implications of our findings, with particular emphasis the existence of Region~II as an inherent limitation of indirect DM detection methods that use signatures of captured DM heating inside astrophysical objects.

\subsection{Closed-form DM capture rates}\label{ssec:SCMS}

In the previous section we reviewed the simple approximations for the total capture rate, valid in four distinct regions of the $\sigma-m_X$ parameter space. Those regions, and the corresponding Eqns.~\ref{eq:CtotRI}-\ref{eq:CtotRIV}, emerge from various approximations of a closed sum form of the $C_N$ partial rate coefficients as given in Eq.~\ref{eq:CN-newaprx}, as explained in detail in Appendix B of Ref.~\cite{Ilie:2023exo}. The schematic picture presented in Fig.~\ref{fig:Regions} and the corresponding simple approximations for the caopture rates presented in Eqns.\ref{eq:CtotRI}-\ref{eq:CtotRIV} were validated numerically in Refs.~\cite{Ilie:2020PopIII,Ilie:2023exo}. Motivated by the fact that, for the case of $\alpha>1$ (i.e. exoplanets or other light capturing bodies), there is a region of the $\sigma-m_X$ parameter space for which no closed form approximation of the total capture rate is found (i.e. white gap region of Fig.~\ref{fig:Regions}) 
in this subsection we derive the most general closed form solution for the total capture rate. Albeit cumbersome, the expression we will present here can be very useful, as it would allow one to bypass complex numerical calculations which in some cases are prone to convergence issues. 

Our starting point would be the most general form of $C_N$ (see Eq.~\ref{eq:CN}), which can be recast as: 

\be\label{eq:CNGeneralA}
C_{N}=A\,p_N(\tau)\left(\left(2 \bar{v}^{2}+3 v_{e s c}^{2}\right)-\left(2 \bar{v}^{2}+3 v_{N}^{2}\right) \exp \left(-\frac{3\left(v_{N}^{2}-v_{e s c}^{2}\right)}{2 \bar{v}^{2}}\right)\right), 
\ee
where, for simplicity we used the definition: $A\equiv \frac{1}{3}\pi \Rstar^2 \sqrt{\frac{6}{\pi}} \frac{n_X}{\vbar}$, consistent with the notation from Eqns.~\ref{eq:CtotRI}-\ref{eq:CtotRIV}. Also, we remind the reader that $v_N=\vesc(1-\avg{z}\beta_+)^{-N/2}$ is the velocity of DM after $N$ scatters. Note that we did not start with the expression of $C_N$ from Eqn.~\ref{eq:CN-newaprx}, since that is based on the binomial expansion of $v_N=\vesc(1-\avg{z}\beta_+)^{-N/2}$, which, in turn, is valid only whenever $\avg{z}\beta_+N\ll1$. The breakdown of the binomial expansion approximation is, in fact, the main reason for the emergence of the white gap region in the case of $\alpha<1$. With this in mind, let's proceed to close the following sum:
\be
C_{tot}=A\,\soneinf \,p_N(\tau)\left(\left(2 \bar{v}^{2}+3 v_{e s c}^{2}\right)-\left(2 \bar{v}^{2}+3 v_{N}^{2}\right) \exp \left(-\frac{3\left(v_{N}^{2}-v_{e s c}^{2}\right)}{2 \bar{v}^{2}}\right)\right),
\ee
i.e. the total capture rate with $C_N$ given as in their most general, approximated form (i.e. Eq.~\ref{eq:CNGeneralA}). We next introduce the following notations:
\begin{eqnarray}
    S_1&\equiv&\soneinf p_N(\tau)\label{Eq:S1}\\
    S_2&\equiv&\soneinf p_N(\tau)\expmess\label{Eq:S2}\\
    S_3&\equiv&\soneinf \pn\vn^2\expmess\label{Eq:S3}.
\end{eqnarray}
With those definitions in mind the total capture rate becomes:

\be\label{Eq:CtotFinalS1S2SS3}
C_{tot}=A\left[(2\bar{v}^2 + 3v_{esc}^2)S_1 - 2\bar{v}^2 S_2 -3S_3\right].
\ee

Next we evaluate the simplest one of those three sums, i.e. $S_1$. Throughout this section we will only consider the $\tau\gg1$ limit, i.e. multiscatter capture. For the case of single scatter capture ($\tau\ll1$) the total capture rate is, by definition, just $C_1$. Note that $S_1$ can be recast as:

\be\label{eq:S1Final}
S_1=\sum_{N=0}^{\infty}\pn-p_0(\tau)\approx 1-\mathcal{O}(1/\tau^2).
\ee
The fact that $\sum_{N=0}^{\infty}\pn=1$ comes from the definition of $\pn$, as the probability of {\it exactly} $N$ scatterings. Moreover,  $p_0(\tau)$ is a $\mathcal{O}(1/\tau^2)$ quantity, as can be seen from the approximation of $\pn$ for $\tau\gg1$ found in Ref.~\cite{Ilie:2020PopIII}:

\be\label{eq:PNapprox}
p_N(\tau)\simeq\frac{2}{\tau^2}(N+1)\Theta(\tau-N).
\ee
In what follows we use the following simplifying notation:
\be\label{eq:r}
r\equiv\frac{3\vesc^2}{2\vbar^2}.
\ee
$S_2$ can now be recast as:

\begin{equation}
    S_2=e^r\soneinf\pn\exp{\left[-r(1-k/r)^{-N}\right]},\label{Eq:S2r}
\end{equation}
Using the approximation of $\pn$ as in Eq.~\ref{eq:PNapprox}  leadds to:

\be
\boxed{
S_2=e^r\frac{2}{\tau^2}\left[\Sigma_1+\Sigma_N\right]\label{Eq:S2Sigmas},
}
\ee
where $\Sigma_1$ and $\Sigma_N$ are the following sums:
\begin{eqnarray}
    \Sigma_1&\equiv&\sum_{N=1}^{\tau}\exp\left[-r(1-k/r)^{-N}\right],\label{eq:Sigma1def}\\
    \Sigma_N&\equiv&\sum_{N=1}^{\tau}N\exp\left[-r(1-k/r)^{-N}\right]\label{eq:SigmaNDef}.
\end{eqnarray}
 We define $B\equiv1-k/r$. Note that this parameter is in the following interval: $B\in[1-\avg{z},1]\subset[0,1]$, since $k/r=\avg{z}\beta_+$, and both $\beta_+$ and $\avg{z}$ are in the $[0,1]$ interval. As shown in Appendix~\ref{Ap:Ctot}, the two sums above can be closed, with the following expressions:

\begin{empheq}[box=\fbox]{align}
    \Sigma_1&\approx-\frac{\Gamma\left(0,rB^{-1},rB^{-\tau}\right)}{\ln{B}}\label{eq:Sigma1},\\
    \Sigma_N&\approx\frac{1}{\ln^2{B}}\left[I\left(rB^{-1},rB^{-\tau}\right)-\ln{r}\Gamma\left(0,rB^{-1},rB^{-\tau}\right)\right].\label{eq:SigmaN}
\end{empheq}
Above,  the function $\Gamma(a,z_l,z_u)\equiv\int_{z_l}^{z_u} t^{a-1}e^{-t}\,dt$ represents the generalized incomplete Gamma function. We also defined $I(a,b)$ as the following definite integral: 
\be\label{eq:Idef}
I(a,b)\equiv\int_a^b\,t^{-1}\ln{t}\exp{(-t)}\,dt,
\ee
For which we find the following closed form solution:
\be\label{eq:IMejier}
\boxed{
I(a,b)=G_{2,3}^{3,0}\left(a\left|
\begin{array}{c}
 1,1 \\
 0,0,0 \\
\end{array}
\right.\right)-G_{2,3}^{3,0}\left(b\left|
\begin{array}{c}
 1,1 \\
 0,0,0 \\
\end{array}
\right.\right)+\ln (a) \Gamma (0,a)-\ln (b) \Gamma (0,b),}
\ee
where by G is the Meijer G-function, and $\Gamma(,)$ stands for the incomplete gamma function. Therefore, the second sum, $S_2$ (see Eq.~\ref{Eq:S2}) is now closed. Using Eqns.~\ref{Eq:S2Sigmas}-\ref{eq:Sigma1}-\ref{eq:SigmaN} we find, after some simplifications:  
\be\label{eq:S2Final}
S_2=\frac{2e^r}{\tau^2}\frac{G_{2,3}^{3,0}\left(\frac{r}{B}\left|
\begin{array}{c}
 1,1 \\
 0,0,0 \\
\end{array}
\right.\right)-G_{2,3}^{3,0}\left(B^{-\tau } r\left|
\begin{array}{c}
 1,1 \\
 0,0,0 \\
\end{array}
\right.\right)+\ln{B} \left[2 \text{Ei}\left(\text{-}\frac{r}{B}\right)-(\tau +1) \text{Ei}\left(\text{-} \frac{r}{B^\tau}\right)\right]}{\ln^2B},
\ee
where $\text{Ei}(z)$ stands for the exponential integral function, defined as: 
$$\text{Ei}(z)\equiv-\int_{-z}^{\infty}\frac{e^{-t}}{t}\,dt$$ 
We next find a closed form for the third sum ($S_3$, defined by Eq.~\ref{Eq:S3}). In terms of the newly defined, {\it{independent}} parameters, $r$ and $B\equiv1-k/r=1-\avg{z}\beta_+$ this sum becomes:

\be\label{eq:S3rB}
S_3=\vesc^2e^r\soneinf\pn B^{-N} \exp{\left[-rB^{-N}\right]}.
\ee
Now, using the form of $S_2$ from Eq.~\ref{Eq:S2r}, and the definition of $B$ one can show that $S_3$ can be simply expressed as:
\be\label{eq:S3Boxed}
\boxed{
S_3=\vesc^2\left[S_2-\frac{\partial S_2}{\partial r}\right].}
\ee
With $S_2$ in terms of $\Sigma_1$ and $\Sigma_N$, as in Eq.~\ref{Eq:S2Sigmas} one gets:

\be\label{eq:S3Sigmas}
S_3=-\vesc^2\frac{2e^r}{\tau^2}\left[\frac{\partial\Sigma_1}{\partial r}+\frac{\partial\Sigma_N}{\partial r}\right].
\ee
Next we use the closed forms presented in Eqns.~\ref{eq:Sigma1}-\ref{eq:SigmaN} to estimate the partial derivatives with respect to $r$ of $\Sigma_1$ and $\Sigma_N$:
\begin{align}
    \frac{\partial\Sigma_1}{\partial r}&=\frac{1}{r\ln{B}}\left[e^{-r/B}-e^{-r/B^{\tau}}\right], \label{eq:partialrSigma1}\\
    \frac{\partial\Sigma_N}{\partial r}&=\frac{e^{-r B^{-\tau }} \ln \left(B^{-\tau }\right)+e^{-\frac{r}{B}} \ln (B)+\Gamma \left(0,B^{-\tau } r\right)+\text{Ei}\left(-\frac{r}{B}\right)}{r \ln ^2(B)}.\label{eq:partialrSigmaN} 
\end{align}
After some simplifications, the final form of $S_3$ becomes:
\be\label{eq:S3final}
\boxed{
S_3=\frac{4 e^r \vbar^2 \left[\Gamma \left(0,rB^{-1},rB^{-\tau}\right)-\ln (B)\left(2 e^{-r B^{-1}}-(\tau +1) e^{-rB^{-\tau}}\right)\right]}{3 \tau ^2 \ln ^2(B)}
}
\ee

Therefore we now have all the pieces in place to get a general closed form of the total capture rate. Specifically, the form of $S_1$, $S_2$ and $S_3$ as in Eqns.~\ref{eq:S1Final}-\ref{eq:S2Final}-\ref{eq:S3final} can be now used into the $C_{tot}$ expressed as in Eqn.~\ref{Eq:CtotFinalS1S2SS3}. In summary, we found the most general analytic form for the capture rate in the multiscatter regime. We did so by starting from the capture rate coefficients $C_N$ expressed as in Eq.~\ref{eq:CN}, and making only one assumption: $\tau\gg1$ (i.e. multiscatter capture). While somewhat cumbersome, it should be useful for bypassing numerical implementations of $C_{tot}$, which are notoriously prone to convergence issues. Moreover, since our simple approximations for $C_{tot}$ (see Eqns.~\ref{eq:CtotRI}-\ref{eq:CtotRIV}) don't cover the entire $\sigma-m_X$ parameters space whenever $\alpha<1$ (see decoupling of RI and RII in right panel of Fig.~\ref{fig:Regions}), we find the expresson for $C_{tot}$ derived here particularly useful when working with objects with $\alpha<1$, such as exoplanets. 

\section{Implications for future work}\label{sec:Implications}

In the previous section we derived the most general form of the DM capture rate (Eqn.~\ref{Eq:CtotFinalS1S2SS3}, supplemented by Eqns.~\ref{eq:S1Final}-\ref{eq:S2Final}-\ref{eq:S3final}), and reviewed our previous simple approximations (Eqns.~\ref{eq:CtotRI}-\ref{eq:CtotRIV}), leading to a four region picture in the $\sigma-m_X$ parameter space (see Fig.~\ref{fig:Regions}). In addition to allowing one to gain important intuition regarding the Dark Matter capture rates in various regimes, our analytic work is quite useful in other ways. For instance, one can quickly check if numerical implementations of the infinite sum defining the total capture leads to expected (or sensible) results. Additionally, as already mentioned, one could simply avoid numerical calculations of $C_{tot}$, and instead use the general analytic form we presented in  this paper, which only assumes: i) a MB distribution for the DM particles and ii) the capturing object at rest with respect to the DM halo. For the more general case of objects placed outside of the immediate vicinity of the center of the DM halo, i.e. with non-zero relative velocity w.r.t. the DM restframe, one can still use the analytic results presented here, and suppress the total capture rate by the analytic factors found by us in Ref.~\cite{Ilie:2021vel}. Lastly, we comment on the possibility of capture via collisions with multiple kinds of targets inside an object. In the future we plan to extend the results presented here in an attempt to provide a analytic closed form for the DM capture rates when including more than one target nucleus, i.e. for the multi-component scattering formalism developed by us in Ref.~\cite{Ilie:2021mcms}. For now, the case of multi-component scattering, to our knowledge, can only be treated numerically.   

Below we discuss the implications of the existence of a region where the capture rate saturates, and becomes $\sigma$ independent (Region~II, i.e. RII). The DM captured inside any object can annihilate and produce detectable signatures, i.e. heat the object up. The ``extra'' heating deposited inside by DM annihilations will be proportional to the capture rate, once there is an equilibrium between capture and annihilations:\footnote{This equilibrium  is reached on relatively short timescales, depending on the annihilation cross section, of course}
\be\label{Eq:LDM}
L_{DM}=m_X C_{tot}
\ee
As such, the heating signal from captured DM annihilations itself is bound to saturate in RII. Thus, if one finds an object in which the extra heating signature is above the saturation limit discussed above there must, necessarily be additional external (or unaccounted for internal) heating sources. At that stage a disambiguation between those two (or more) ``extra'' sources of heating (DM annihilations and something else) is practically impossible. Therefore, the method described above to probe DM-nucleon interactions with astrophysical probes loses all of its constraining power in RII of the $\sigma$ vs. $m_X$ parameter space. Conversely, that implies that if they are hotter than a certain critical temperature ($T_{crit}$) those objects cannot be used as DM probes.  We will next give several examples to make this discussion more concrete.   

Old neutron stars are particularly good DM probes, especially in the high $m_X$ regime as shown by~\cite{Bramante:2017}. This is due to them being extremely dense, and as such, amazingly efficient DM captors. In~\cite{Ilie:2020Comment} we find (see Eq.~17 in~\cite{Ilie:2020Comment}) the critical temperature for a NS of mass $M_{NS}$ and radius $R_{NS}$ placed inside a DM environment with density $\rho_X$ and a dispersion velocity $\bar{v}$:

\be\label{eq:TcritNS}
T_{crit}= 2.04 \times 10^{4} ~\unit{K} \left(1-\frac{r_S}{R_{NS}}\right)^{-1/4}\left(\frac{R_{10}}{R_{NS}}\frac{M_{NS}}{M_{1.5}}\frac{v_{220}}{\vbar}\frac{\rho_X}{\rho_{1000}}\right).
\ee
We introduced the following simplifying notations: $R_{10}\equiv10$~km, $v_{220}\equiv220$~km/s, $M_{1.5}\equiv1.5~M_{\odot}$, $r_S\equiv 2GM_{NS}/c^2$, and $\rho_{1000}=10^3~\unit{GeV}~\unit{cm}^{-3}$. For instance, for the benchmark parameters given above we get $T_{crit}\approx 24,000~\unit{K}$. Note that $\rho_X=\rho_{1000}$ implies a NS within the inner $\sim 10$~pc of the galactic center (GC). This implies that NS within the GC vicinity cannot be used as DM probes if they are hotter than $T_{crit}\simeq24,000$~K. Which, in turn could limit the detectability of such probes.  

Exoplanets and brown dwarfs could serve as excellent probes of DM with sub-GeV particle mass, as pointed out by Ref.~\cite{Leane:2020wob}. Assuming those objects (with radius $R$) are in thermal equilibrium and have an observed surface temperature $T_{eff}$ we have, according to the Stephan-Boltzmann law:
\be\label{eq:SB}
\Gamma^{DM}+\Gamma^{ext}+\Gamma^{int}=\Gamma^{rad}=4\pi\sigma_{SB} R^2 T_{eff}^4\epsilon,
\ee
where each of the $\Gamma$'s is the heating rate due to, in turn: the DM annihilations, any possible external sources, and the internal heating, in absence of DM anihilations. Here $\epsilon$ represents the emissivity, a measure of how efficiently the object radiates away its energy. The best DM probes are those for which there is no competing heat source, so, in what follows we assume $\Gamma^{ext}=0$, following~\cite{Leane:2020wob}. Moreover, $\Gamma^{int}$ can be approximated as described below. The best case scenario is an object with the minimum internal heating. That makes exoplanets and brown dwarfs such good DM probes. However, there is a lower limit for $\Gamma^{int}$ that is set by the minimum surface temperature they can attain (in the absence of any other heat sources) at a given age ($T_{min}$). The older they are, the cooler they would be, and for Milky-Way probes a natural cutoff in the age is around $10$~Gyrs. From simulations one can estimate $T_{min}$ as a function of age, for a given object. For instance for heavy BDs (i.e. $M\simeq 55M_{jup}$) one has $T_{min}\approx750$~K, whereas for a benchmark Jupiter-like planet ($M=M_{jup}$ and $R=R_{jup}$) one has $T_{min}\approx 80$~K. Now, the fact that $C_{tot}$ saturates in RII implies, as discussed, that $\Gamma^{DM}=m_X C_{tot}$ saturates to $m_X C^{II}$. Thus, one finds $T_{crit}$ as a solution to the following:
\be\label{eq:TcritConditionEP}
m_X C^{II}=4\pi\sigma_{SB} R^2 \left(T_{crit}^4-T_{min}^4\right)\epsilon
\ee
This can be solved as:
\begin{equation}\label{Eq:TcritEp}
    T_{crit} = \left( \frac{1}{\sqrt{24\pi}} \frac{\rho_X}{\sigma_{SB} \epsilon\vbar} \left(2\vbar^2 + 3 v_e^2\right) + T_{min}^4\right)^{\frac{1}{4}}.
\end{equation}
Since $T_{crit}$ increases with $\rho_X$ and $1/\epsilon$, the most favorable scenario is a high $\rho_X$ and(or) a low $\epsilon$. We first analyze the $\epsilon=1$ case. Taking $\rho_X\simeq 10^3~\GeV\percc$ we get $T_{crit}\simeq 760$~K for heavy brown dwarfs ($M=55 M_{jup}$ and $R=R_{jup}$). This leaves a very very narrow window of temperatures available in which those objects can serve as DM probes: from 750K (i.e. $T_{min}$) to 760K (i.e. $T_{crit}$). For the benchmark Jupiter-like planets we get a $T_{crit}\simeq 156$~K, which is falls short of the minimum temperature need for detection near the galactic center with JWST, which, according to Ref.~\cite{Leane:2020wob}, is $650$~K. This therefore rules out lowest mass Jupters as DM probes, whenever their emissivity is close to unity. By setting $T_{crit}=650$~K we can find, via Eq.~\ref{Eq:TcritEp}, we find that $\sim M_J$ is the minimum mass a Jupiter-like planet placed in the GC vicinity has to have in order to be detectable while also having any chance of being useful as an exoplanet. The situation improves when we consider lower emissivity, as expected. For instance, assuming $\epsilon=0.001$ we now get $T_{crit}\sim2000$~K for the heavy BDs and $T_{crit}\simeq 860$~K for the benchmark Jupiter-like planets with mass $M=M_{jup}$. 

In summary, in order for any object to be useful as a DM probe its observed temperature must be lower than the critical temperature ($T_{crit}$). For Neutron Stars, or other compact objects for which general relativistic effects are important, $T_{crit}$ can be estimated according to Eq.~\ref{eq:TcritNS}. For all other objects $T_{crit}$ can be found according to Eq.~\ref{Eq:TcritEp}.

\appendix
\section{Closed form for the two sums labeled $\Sigma_1$ and $\Sigma_N$}\label{Ap:Ctot}

In this appendix we show that indeed $\Sigma_1$ and $\Sigma_N$, the two sums defined as in Eqns.~\ref{eq:Sigma1def}-\ref{eq:SigmaNDef} can be closed, as in Eqns.\ref{eq:Sigma1}-\ref{eq:SigmaN}, whenever $\tau\gg1$, i.e. in the multiscatter regime. In this case the sums are good approximations to the following integrals:
\begin{align}
    \Sigma_1&\approx\int_1^\tau\exp{(-r B^{-N})}\, dN\\
    \Sigma_N&\approx\int_1^\tau N\exp{(-r B^{-N})}\, dN.
\end{align}
Changing variables to $t\equiv rB^{-N}$ it is straightforward to show that $dN=-\frac{1}{\ln{B}}\frac{dt}{t}$, and therefore:
\begin{align}
    \Sigma_1&\approx-\frac{1}{\ln{B}}\int_{r/B}^{r/B^{\tau}}\frac{e^{-t}}{t}\, dt\label{eq:Sigma1t}\\
    \Sigma_N&\approx\frac{1}{\ln^2{B}}\int_{r/B}^{r/B^{\tau}}\frac{e^{-t}}{t}\ln{\frac{t}{r}}\, dt\label{eq:SigmaNt}.
\end{align}
The first integral, appearing in Eq.~\ref{eq:Sigma1t} can be identified as the incomplete gamma function $\Gamma(0,r/B,r/B^\tau)$, thus proving $\Sigma_1$ as in Eq.~\ref{eq:Sigma1}. In order to get $\Sigma_N$ from Eq.~\ref{eq:SigmaN}, we note that $\ln{t/r}=\ln{t}-\ln{r}$.

\bibliographystyle{JHEP}
\bibliography{RefsDM}
\end{document}